\documentclass[prl,amsmath,amssymb,superscriptaddress,twocolumn]{revtex4-2}

\usepackage{graphicx}
\usepackage{dcolumn}
\usepackage{epsfig}
\usepackage{float}

\usepackage[bbgreekl]{mathbbol}

\usepackage{mathrsfs}

\usepackage[cal=boondox,scr=boondoxo]{mathalfa}

\usepackage{trsym} 

\newcommand\vex[1]{\mathbf{#1}}
\newcommand\gvex[1]{\boldsymbol{#1}}

\usepackage[T3,T1]{fontenc}
\DeclareSymbolFont{tipa}{T3}{cmr}{m}{n}
\DeclareMathAccent{\invbreve}{\mathalpha}{tipa}{16}

\usepackage{accents}
\newlength{\hhatheight}

\usepackage{xcolor}

\makeatother

\begin{document}

\title{Dirac Magic and Lifshitz Transitions in AA-Stacked Twisted Multilayer Graphene}

\author{Yantao Li}
\affiliation{Department of Physics, Indiana University, Bloomington, Indiana 47405, USA}

\author{Adam Eaton}
\affiliation{Department of Physics, Indiana University, Bloomington, Indiana 47405, USA}

\author{H. A. Fertig}
\affiliation{Department of Physics, Indiana University, Bloomington, Indiana 47405, USA}
\affiliation{Quantum Science and Engineering Center, Indiana University, Bloomington, Indiana 47405, USA}

\author{Babak Seradjeh}
\affiliation{Department of Physics, Indiana University, Bloomington, Indiana 47405, USA}
\affiliation{Quantum Science and Engineering Center, Indiana University, Bloomington, Indiana 47405, USA}
\affiliation{IU Center for Spacetime Symmetries, Indiana University, Bloomington, Indiana 47405, USA}

\begin{abstract}
We uncover a new type of magic-angle phenomena when an AA-stacked graphene bilayer is twisted relative to another graphene system with band touching. In the simplest case this constitutes a trilayer system formed by an AA-stacked bilayer twisted relative to a single layer of graphene. We find multiple  anisotropic Dirac cones coexisting in such twisted multilayer structures at certain angles, which we call ``Dirac magic.'' We trace the origin of Dirac magic angles to the geometric structure of the twisted AA-bilayer Dirac cones relative to the other band-touching spectrum in the moir\'e reciprocal lattice. The anisotropy of the Dirac cones and a concomitant cascade of saddle points induce a series of topological Lifshitz transitions that can be tuned by the twist angle and perpendicular electric field. We discuss the possibility of direct observation of Dirac magic as well as its consequences for the correlated states of electrons in this moir\'e system.
\end{abstract}

\date{\today}

{
\let\clearpage\relax
\maketitle
}

\emph{Introduction}.---%
The discovery of correlated electronic states in twisted bilayer graphene has ushered in a new era of ``twistronics'' in stabilizing novel phases of quantum matter in low dimensions~\cite{dosSantos_2007,Shallcross_2008,Shallcross_2010,Mele_2010,Mele_2011,Bistritzer_2011,dosSantos_2012,Carr_2017,cao2018correlated,cao2018unconventional,Tarnopolsky_2019,yankowitz2019tuning}. A growing number of twisted structures of layered van der Waals materials, such as multilayer graphene~\cite{Koshino_2019,Khalaf_2019,Liu_2019,cao2021large,Tritsaris_2020,Guerci_2021,Fischer_2021,Topp_2019,Li_2020,Katz_2020,Assi_2021} and transition metal dichalcogenides~\cite{Fengcheng_2018,Naik_2018,Tang2020,Tran_2020,Zhang_2020}, have been studied theoretically and experimentally. Some of these systems exhibit a series of  ``magic angles,'' characterized by low-lying bands of bandwidth much smaller than the energy scales of the original layers and their tunnel coupling. Interactions play a prominent role in determining the electronic state when the Fermi energy is in such a band~\cite{Kang_2018,Po_2018,Xu_2018,Kang_2019}.

The appearance of flat bands can be understood intuitively from the competition between the dispersive energy scale of each layer, e.g.~the Dirac cones in graphene, and the tunneling energy scale modulated by the moir\'e pattern of twisted strcuture. At small twist angles, the latter yields a nearly periodic moir\'e superlattice or, equivalently, a small moir\'e Brillouin zone repeated in the reciprocal space. Magic angles are found when the energy of the twisted bands at successive moir\'e Brillouin zones become comparable to the tunneling energy.
This picture raises the question of whether other interesting band reconstructions can arise from competing energy scales. In particular, a heterogeneous choice of the
twisted layers broadens the possibilities for twistronics~\cite{Spanton_2018,Shabani_2021,Zhou_2021}.

In this paper, we uncover a new type of twisted band engineering in multilayer graphene, formed by an AA-stacked graphene bilayer twisted relative to another graphene layer with degenerate band touchings (D), such as a single sheet of graphene with a Dirac cone in the simplest case, Bernal-stacked graphene with quadratic band touching~\cite{Jung_2014}, or rhombohedral graphene trilayer with cubic band touching~\cite{Zhang_2010,zhou2021superconductivity}. The choice of AA stacking for the bilayer is motivated by its close relation to the degenerate graphene layer, effectively consisting of two Dirac cones shifted to finite energies by the tunneling amplitude between the layers. We find that this geometry hosts special angles for which multiple Dirac cones coincide {at the same high symmetry points in a moir{\'e} Brillouin zone}. We call this phenomenon ``Dirac magic.''

For an AA-stacked graphene bilayer twisted on a single sheet of graphene, we find that on approaching Dirac magic angles the energy bands near the Dirac point of the single layer undergo significant reconstruction. This is a result of a topological transition where the local maximum of the second band becomes a local minimum as $C_3$-symmetric saddle points merge. We find that this process generates a cascade of saddle points that spiral toward the Dirac point. As a result, a series of topological Lifshitz transitions are induced {by varying the Fermi level near neutrality~\cite{Carter_2018,Hejazi_2019}.} We show that in addition to variations in the twist angle, these Lifshitz transitions can be tuned by a perpendicular electric field. Thus, the Dirac magic phenomena enable new types of twisted band engineering and provide a rich platform for correlated electronic states.

\emph{Model}.---%
The general form of the Hamiltonian for the system we study is
\begin{equation}
H_\text{AA/D} = \begin{bmatrix} h_{\text{AA},\theta/2} & T \\ T^\dagger & h_{\text{D},-\theta/2} \end{bmatrix},
\end{equation}
where $h_{\text{L},\pm\theta/2}$ is the Hamiltonian of layer $\text{L}=\text{AA, D}$ at twist angle $\pm\theta/2$ and $T$ is the tunneling between the adjacent sheets of the layers. For concreteness, we consider the simplest case of an AA-stacked bilayer twisted relative to a single layer, see Fig.~\ref{fig:sketch}(a). Results for AA/AB and AA/ABC twisted graphene multilayers are presented in the Supplemental Material~\footnote{See Supplemental Material for more details of the band spectra at Dirac magic angles of AA/S, AA/AB, and AA/ABC twisted heterostructures.}. Following Bistritzer and MacDonald~\cite{Bistritzer_2011}, we model the trilayer system with
\begin{equation}
H_\text{AA/S}=
\begin{bmatrix}
h_{\theta/2} - V & T_\text{AA} & 0 \\
T_\text{AA}^\dagger & h_{\theta/2} & T\\
0 & T^\dagger & h_{-\theta/2} + V \\
\end{bmatrix},\label{eq:AAS}
\end{equation}
where $h_{\theta/2} = - i\hbar v_F \gvex\nabla \cdot\gvex\sigma_{\theta/2}$ is the low-energy single-layer Hamiltonian with Fermi velocity $v_F$ and rotated Pauli matrices $\gvex\sigma_{\theta} \equiv R_\theta^\dagger (\sigma_x,\sigma_y) R_\theta$, where $R_{\theta} = e^{i\theta\sigma_z/2}$. Within the AA-bilayer we take
\begin{equation}
T_\text{AA}=
\gamma_\text{AA} + i \gamma_\text{tw} \gvex\nabla\cdot\gvex\sigma_{\theta/2},
\label{eq:TAA}
\end{equation}
where $\gamma_\text{AA}$ is onsite tunneling and $\gamma_\text{tw}$ the trigonal warping of the bilayer bands. The tunneling matrix between the single layer and one of the AA-stacked sheets is $T = \sum_{n=1}^3 T_n e^{-i k_\theta \vex q_n \cdot \vex r}$,
\begin{equation}\label{eq:T}
T_n = w \left( u + \vex q_n \cdot \gvex\sigma_{\pi/2} \right),
\end{equation}
where $\vex q_1 = (0,-1)$ and $\vex q_{2,3} =  (\pm\sqrt3/2,1/2)$ are the wavevectors associated with the moir\'e superlattice in units of $k_\theta = 8\pi \sin(\theta/2)/3 a$, $a$ is the Bravais lattice spacing of graphene, $w$ and $uw$ are the tunneling amplitudes between the AA- and AB-regions of the moir\'e pattern. We have also included the potential bias $V$ to model an electric field perpendicular to the layers. In our numerical results below we take $a= 2.4$~\AA, $\hbar v_F / a = 2.425$~eV, $\gamma_{AA}=217$~meV, $\gamma_\text{tw}=20$~meV, $w = 110$~meV and $u = 0.816$ unless otherwise noted~\cite{Lobato_2011}.

\emph{Geometric origin of Dirac magic}.---%
To understand the origin of Dirac magic angles, we will use a perturbative scheme in the moir\'e Brillouin zone as a function of $w$. We note that in the absence of tunneling ($w=0$), and {neglecting trigonal warping ($\gamma_\text{tw} \rightarrow 0$)}, the moir\'e pattern gives rise to folded bands with zero-energy states at the $K$ point of the single layer and at circles of radius $k_0\equiv\gamma_\text{AA}/\hbar v_F$ centered at $K'$ point of shifted Dirac cones of the bilayer. As shown in Fig.~\ref{fig:sketch}(b), upon lowering the twist angle, these circles first pass through the $K$ point when $k_0=k_{\theta_1}$ for $\theta_1 = 2\sin^{-1}(3a \gamma_\text{AA}/8\pi \hbar v_F) = 1.22^\circ$. As illustrated in Fig.~\ref{fig:Dirac}(a), this is the first of a series of angles at which the zero-energy circles centered at $K'$ in higher-order moir\'e Brillouin zones intersect at the $K$ point, satisfying $k_0 = r_n k_{\theta_n}$ for a series of ratios $r_n = 1, 2, \sqrt 7, \sqrt{13}, 4, \sqrt{19}, \cdots$. Since the angles are small, we find $\theta_n \approx \theta_1/r_n$.

For small $w\neq 0$, the states of the bilayer and single layer mix and, generically, split away from zero energy governed by symmetry. Since the Hamiltonian~(\ref{eq:AAS}) at the $K$ point has three-fold rotational symmetry, $C_3=\text{diag}(R_{2\pi/3},R_{2\pi/3},R_{2\pi/3}) \mathcal{R}_{2\pi/3}$ with $\mathcal{R}_\theta$ the spatial rotation by $\theta$ around $z$ axis, the mixing occurs within each eigenvalue sector of $\text{Spec}(C_3) = \{1, \varphi, \varphi^{-1}\}$, $\varphi = e^{2\pi i /3}$.

\begin{figure}[t]
   \centering
   \includegraphics[width=3.45in]{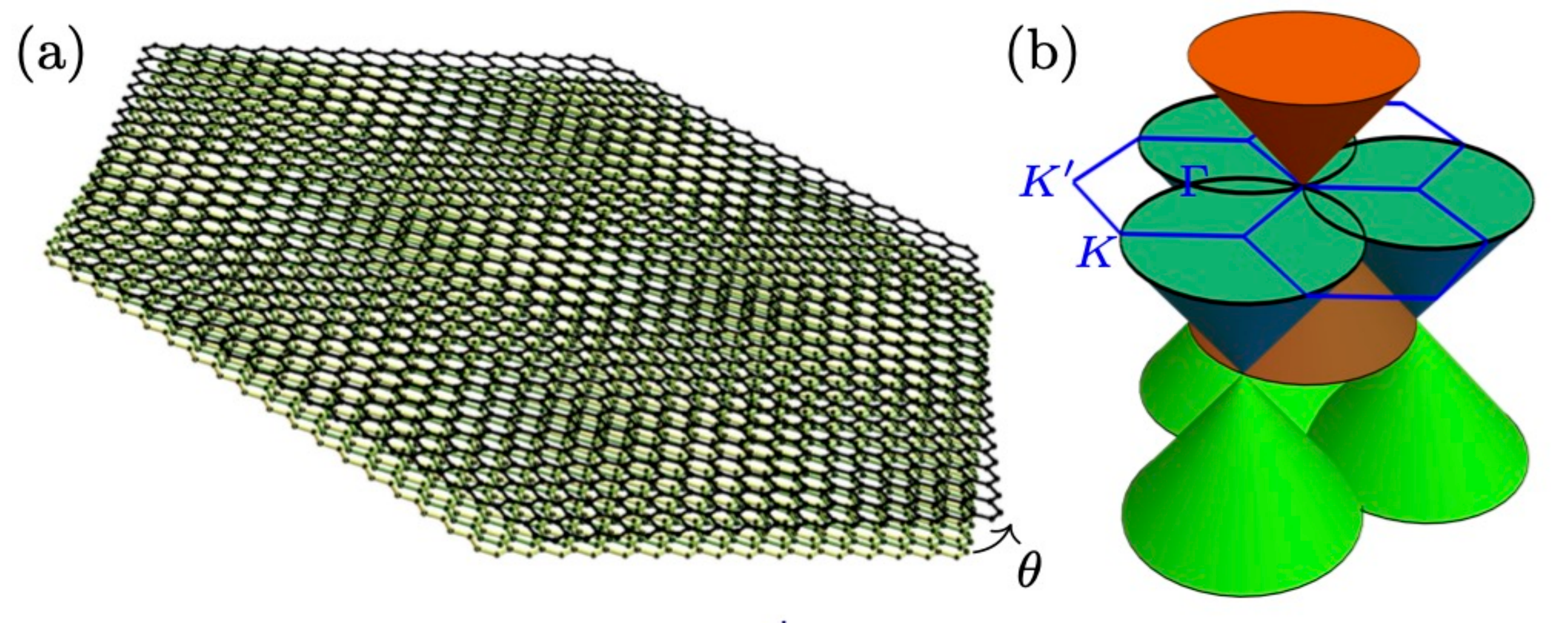} 
   \caption{Sketch of (a) the real-space geometry of the AA/S twisted trilayer heterostructure and (b) the momentum-space energy bands near the Dirac point of the single layer. In (b) the three negative-energy cones of the AA-stacked bilayer and the moir\'e Brillouin zones are shown at the matching condition that originates the first Dirac magic angle.}
   \label{fig:sketch}
\end{figure}

For $u=0$ the Hamiltonian is also chirally symmetric $\{H,C\} = 0$ with $C = \text{diag}(-\sigma_z,\sigma_z,\sigma_z)$ for all $w$, which restricts mixing within opposite chiral eigenvalues $\pm1$. Since $[C,C_3]=0$, the zero-energy states can be taken to be simultaneous eigenstates of $C_3$ and $C$: the three degenerate pairs of states of the AA-bilayer at the $K$ point take all distinct eigenvalues of $C_3$ and $C$, while those of the single sheet have the pair of eigenvalues $(\varphi,+1)$ and $(\varphi^{-1},-1)$, respectively, of ($C_3$, $C$)~\cite{Herbut_2009,Po_2018}. Thus, each $C_3$ sector $\varphi$ or $\varphi^{-1}$ has three states, whose mixing will yield one state at $E_1 = 0$ and a pair of states at energies $\pm E_3\propto w$. All of these energies are doubly degenerate due to the mirror symmetry $M_1$ with mirror plane $K$-$\Gamma$, which satisfies $M_1C_3M_1 = C_3^{-1}$ and maps the $\varphi$ and $\varphi^{-1}$ sectors to one another.

The pair of states in the $C_3$ sector with eigenvalue $1$ are not protected against splitting by tunneling; however, since they both reside in the AA-bilayer for $w=0$, they can only split by mixing with higher-energy states of the single sheet. Thus their splitting results in energies $\pm E_2 \propto w^2$. Indeed, this splitting can be made very small by adjusting the twist angle $\theta$.

\emph{Numerics}.---%
We now demonstrate the physics described above by numerically diagonalizing the Hamiltonian in Eq.~(\ref{eq:AAS}) in a plane wave basis, using the reciprocal lattice formed by $\vex q_2 - \vex q_1$ and $\vex q_3-\vex q_1$. To achieve convergence as the twist angle is lowered, we organize the increasing number of wavevectors  by a given number, $s$, of tunneling matrix elements to produce perturbative results up to $\mathcal{O}(w^s)$ as shown in Fig.~\ref{fig:Dirac}(a). We plot the calculated energy difference $E_2-E_1\equiv\delta E_K(w,\theta)$ at the $K$ point for $u=0.816$ in Fig.~\ref{fig:Dirac}(b). The loci of $\delta E_K$ minima converge to the geometric values $\theta_n$ as $w\to 0$ and reveal the evolution of the Dirac magic angles for $w>0$.

We find the minima of $\delta E_K \approx 0.1$-$0.2$~meV are three orders of magnitude smaller than the tunneling energy scale $w\sim 100$~meV. Moreover, for $w>0$ the tunnel coupling of the zero-energy circles of the AA-bilayer breaks them up into arcs passing through the $K$ point, resulting in a highly anisotropic dispersion of nearly degenerate Dirac cones at Dirac magic angles. Thus, we expect a rich spectral topology at and near Dirac magic angles that can be tuned by the twist angle as well as external fields.

\begin{figure}[t]
   \centering
   \includegraphics[width=3.4in]{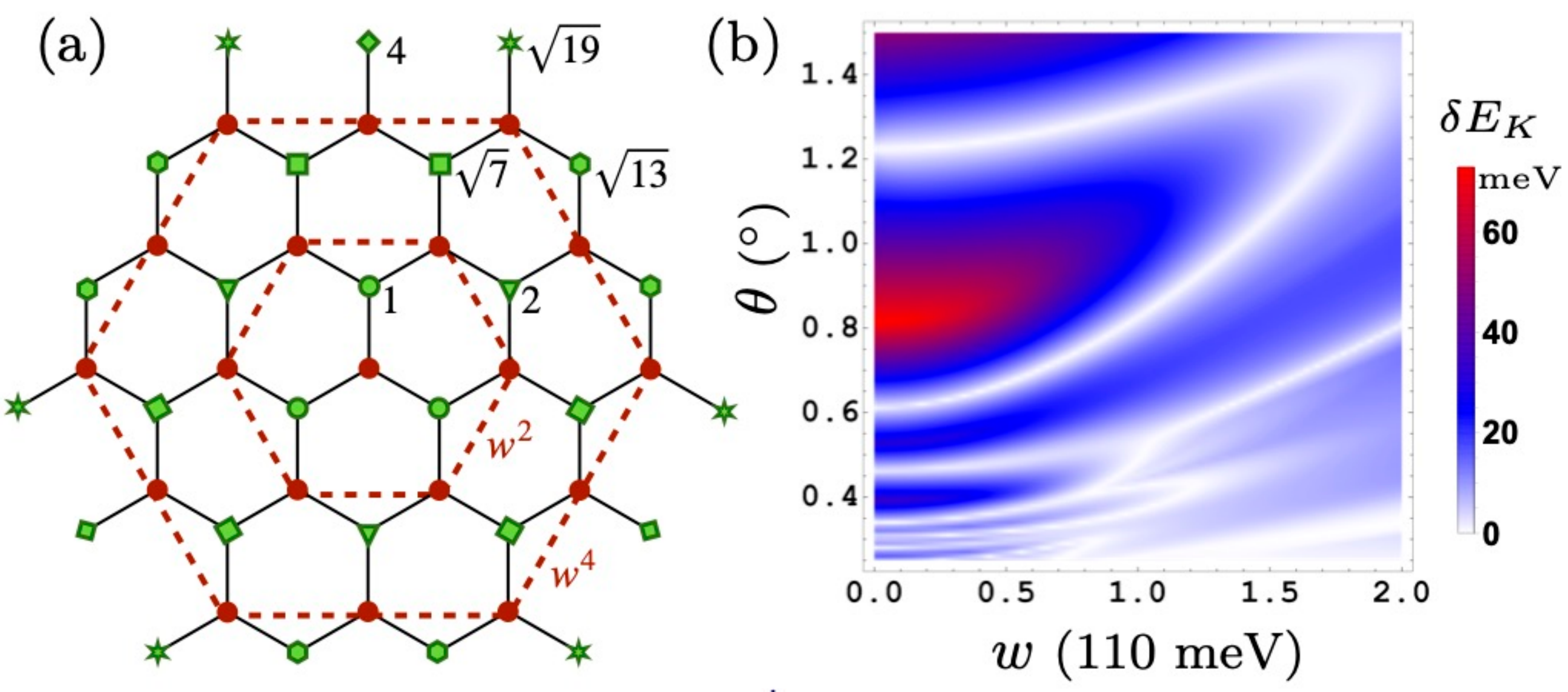} 
   \caption{(a) Reciprocal lattice of the moir\'e superlattice with the $K$ point (red) at the origin and $K'$ points (green) at a distance $r_n k_\theta$ for $r_n=1, 2, \sqrt{7}, \sqrt{13}, 4, \sqrt{19}$ (circle, triangle, square, hexagon, diamond, star). The dashed lines mark the $K$ points connected to the origin at  a given $\mathcal{O}(w^s)$ of tunneling matrix elements (solid lines).
(b) The energy splitting $\delta E_K = E_2-E_1$ at $K$ point at $\mathcal{O}(w^{9})$ for $u=0.816$.}
   \label{fig:Dirac}
\end{figure}

\emph{Dirac magic spectra}.---%
We focus on the second Dirac magic angle, $\theta_2=0.826^{\circ}$ for $w=110$~meV and $u=0.816$, since it has a direct gap between positive- and negative-energy bands away from the $K$ point, allowing for clear experimental signatures. 
(See Supplemental Material~\cite{Note1} for the first Dirac magic angle $\theta_1=1.31^{\circ}$.)


The band spectra along symmetry lines are shown in Fig.~\ref{fig:Band}. The near degeneracy of the four central bands at the $K$ point in Fig.~\ref{fig:Band}(a) is accompanied by significant anisotropy in the $K$-$M$-$K'$ and $K$-$\Gamma$ directions, and a small flat-band feature with a width of about 1.5~meV in the $K$-$\Gamma$ direction. Interestingly, there is a second band minimum visible along the $K$-$\Gamma$. Thus, changing the electron density away from neutrality (for example by a gate voltage) can change the Fermi surface topology through a Lifshitz transition.

Fermi surface topology can also be tuned by changing the twist angle and/or applying a perpendicular electric field. In Fig.~\ref{fig:Band}(b), we show the band spectra at $\theta = 0.85^\circ$, showing the opening of the second band gap at the $K$ point while the near band-crossing between the first and the second band moves away from the $K$ point and to nonzero energies. Moreover, an additional first band minimum develops along the $K$-$M$-$K'$ direction. Applying a perpendicular electric field as in Figs.~\ref{fig:Band}(c) and (d) also opens a gap for the second band at the $K$ point as well as accentuating the avoided band crossing away from the $K$ point. It also reshapes the band minima, thus providing an additional knob to control the Fermi surface topology.

\begin{figure}[t]
   \centering
   \includegraphics[width=3.4in]{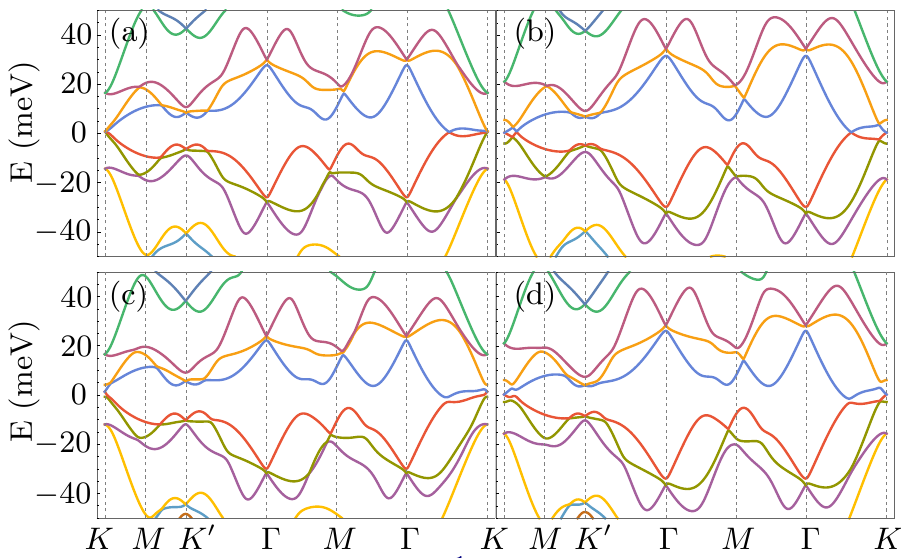} 
   \caption{Band spectra along symmetry lines of the moir\'e Brillouin zone near the second Dirac magic angle for (a) $\theta=0.826^{\circ}, V=0$, (b) $\theta=0.85^{\circ}, V=0$, (c) $\theta=0.826^{\circ}, V= 72.75$~meV, and (d) $\theta=0.85^{\circ}, V= 72.75$ meV. In (a) the minimum gap $\delta E_K = 0.26$~meV. In (b) the gap at band crossings away from the $K$ point is too small for our numerical resolution.}
   \label{fig:Band}
\end{figure}

\begin{figure*}[t]
\begin{center}
\includegraphics[width=6.5in]{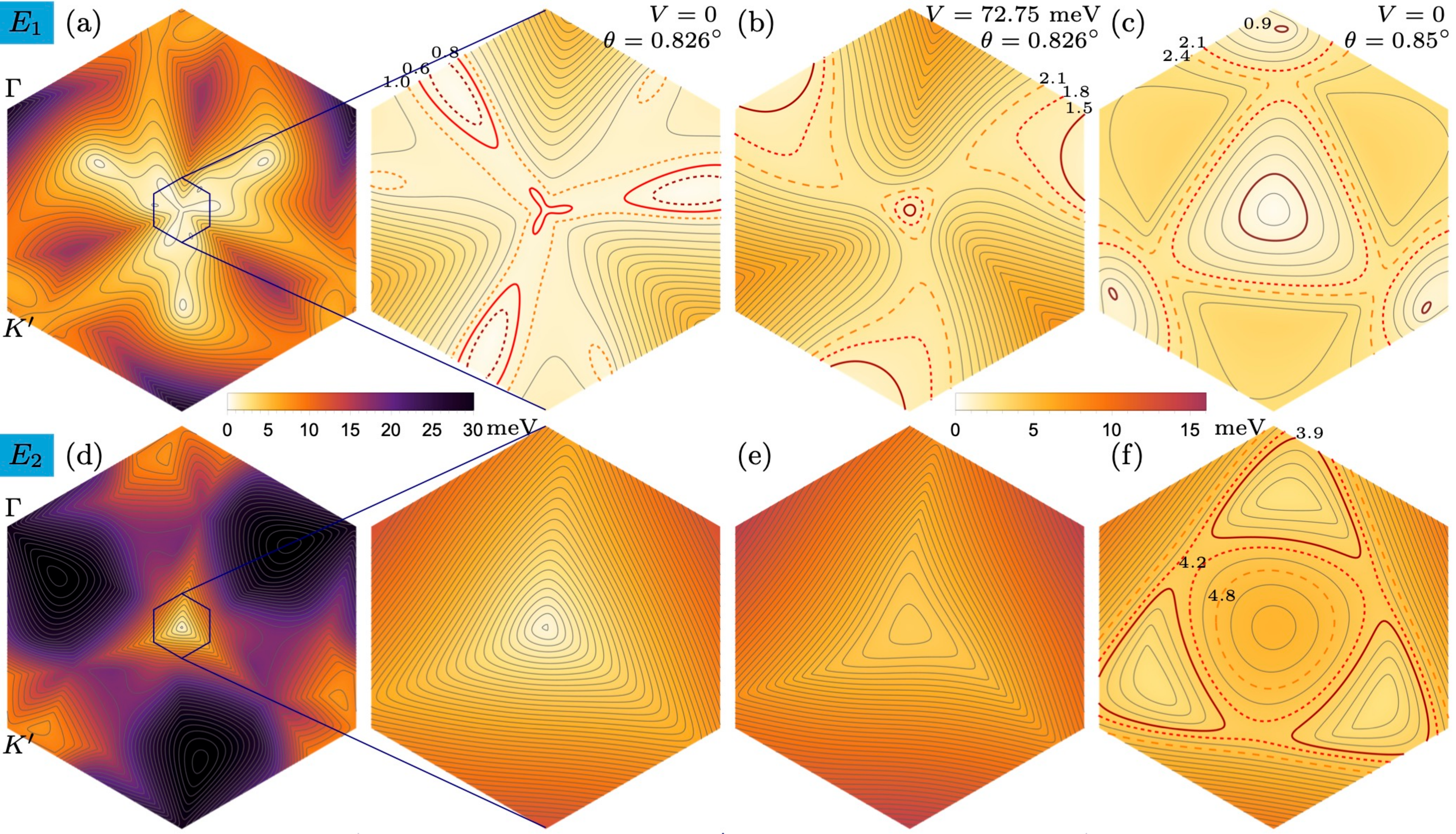}
\caption{Density plot of the first (top row) and second (bottom row) positive bands near the second Dirac magic angle for $w=110$~meV and $u=0.816$. The $K$ point is at the center and the left panels in (a) and (d) show the full moir\'e Brillouin zone with contour lines separated by 1~meV. The rest of the panels are zoomed in around the $K$ point with gray contour lines that are separated by 0.3~meV and a few specifically labeled for illustration. The other parameters are (a), (d) $\theta = 0.826^\circ, V=0$; (b), (e) $\theta=0.826^\circ, V = 72.75$~meV; (c), (f) $\theta = 0.85^\circ, V=0$. The same color bar is used for all plots.
}
\label{fig:DP}
\end{center}
\end{figure*}

\emph{Lifshitz transitions}.---%
To illustrate the Fermi surface topology, we present the topography of the first and second band spectra in the Brillouin zone in Fig~\ref{fig:DP}. This makes the anisotropy in the $K$-$K'$ and $K$-$\Gamma$ directions plainly clear in both bands. In the first band, in particular, a three-blade propeller-shaped flatband region along the $K$-$\Gamma$ direction is evident. Strikingly, as shown by the zoomed panel in Fig~\ref{fig:DP}(a) at Dirac magic angle, this propeller-shaped structure repeats itself at lower energies at least three times in our numerical resolution. This remarkable structure is one of our main findings.

As a consequence of the repeated propeller-shaped structure in this sytem, the approach to the Dirac magic angle at $\theta=0.826^\circ$ is accompanied by a cascade of saddle points. Thus, multiple Lifshitz transitions appear as a function of density or Fermi energy. The second band, by contrast, does not exhibit such an intricate saddle point structure.

Applying a perpendicular electric field reshapes the bands, as shown in Figs.~\ref{fig:DP}(b) and (e). The propeller-shaped contours of constant energy close to the $K$ point become less anisotropic, albeit still triangular. This terminates the cascade of saddle points at low energies. The main effect on the second band is to raise the overall energy scale without introducing saddle points or additional degeneracies with the first band.

Adjusting the angle away from the Dirac magic angle also reshapes the bands; a typical example is shown in Figs.~\ref{fig:DP}(c) and (f). As with the perpendicular electric field, the first band evolves here by merging saddle points and developing a less anisotropic topography around the $K$ point. By contrast, the second band displays a maximum at the $K$ point and a contour of (near) degenerate states with the first band away from it. This is necessarily accompanied by additional saddle points around the $K$ point in the second band. Thus, different Lifshitz transitions appear in higher bands near the $K$ point.

\emph{Discussion}.---%
The appearance of Lifshitz transitions can, in principle, be observed in spectral probes of the bands, e.g. angle-resolved photoemission spectroscopy \cite{Jones_2020} or Landau level spectroscopy \cite{Varlet_2014,Choi_2021}. The change in the Fermi surface topology can also be probed via quantum oscillations. 
Saddle points slow the semi-classical motion of electrons, introducing anomalous drops in the frequency of these oscillations as a function of Fermi energy~\cite{Fertig_1987,Itskovsky_2005,Lu_2014a}. 

A remarkable consequence of the saddle point cascade supported by these AA-twisted systems is the expectation that, with increasing magnetic field (and decreasing magnetic length), increasingly fine structure associated with the Lifshitz transitions should become evident near zero energy, in contrast to single layer graphene, for which only a single degenerate Landau level is present.  Such structure could be revealed via compressibility measurements~\cite{Young_2012}.

The propeller-shaped Fermi surfaces at the Dirac magic angle create quasiparticle scattering wavevectors yielding distinct quasiparticle interference patterns in scanning tunneling spectroscopy~\cite{Crommie_1993,Chen_2017,Zhang_2019}. The nesting of the Fermi surface by these wavevectors could also open the possibility of instabilities toward magnetic and/or charge ordering in the system \cite{Makogon_2011,Long_2016,Sharpe_2019,Fleischmann_2020}.

More generally, the van Hove singularities at the saddle points can promote various correlated electronic phases. The apparent proliferation of such saddle points at low energy near Dirac magic angles points to a potentially rich phase diagram of correlated states at low energies \cite{Sherkunov_2018,Cea_2019,Lake_2021,Xu_2021}. Our study of band topology near Dirac magic angle and its evolution with the twist angle and perpendicular electric field are a crucial first step toward understanding correlated electronic states that can occur in this system.
In this regard, we note that while the existence of Dirac magic angles is a robust feature of the AA-stacked twisted multilayer graphene systems we have studied, details of the band topology they support are sensitive to the values of twist angle and perpendicular electric field. Thus, spatial inhomogeneities in twist angle, local potential imbalance, and electronic densities can result in coexistence of different kinds of order in the same system.

\emph{Summary}.---%
We have introduced a new design concept for twisted moir\'e systems,
in which the geometric matching of a band touching point in one layer and degenerate Fermi surfaces of an AA-stacked graphene bilayer, achieved at certain Dirac magic twist angles, can lead to the appearance of multiple near-degenerate, anisotropic Dirac cones. The approach to these Dirac magic angles is accompanied by a cascade of saddle points and propeller-shaped constant-energy contours in the low-energy bands. This opens the possibility of engineering multiple Dirac cones and tunable Lifshitz transitions in situ. We hope that these findings will stimulate further theoretical and experimental studies of correlated phases in this system.

\begin{acknowledgments}
This work is supported in part by the NSF through the CAREER award DMR-1350663 in early stages, as well as via grant No. DMR-1914451, and ECCS-1936406. Further support was supplied by the US-Israel Binational Science Foundation grant No. 2016130, and the Vice Provost for Research at Indiana University, Bloomington through the Faculty Research Support Program. HAF acknowledges the support of the Research Corporation for Science Advancement through a Cottrell SEED Award.
\end{acknowledgments}

\vspace{-5mm}
\onecolumngrid
\newpage

\renewcommand{\thefigure}{S\arabic{figure}}
\renewcommand{\theequation}{S\arabic{equation}}
\setcounter{equation}{0}
\setcounter{figure}{0}

\title{
Supplemental Material for ``Dirac Magic and Lifshitz Transitions in AA-Stacked Twisted Multilayer Graphene''
}

\begin{abstract}
Here we provide details of Dirac magic and Lifshitz transitions. For the trilayer system, we show the band spectra at the origin of Dirac magic angles, the spectra of the first Dirac magic angle and the first Gamma magic angle. To see more clearly the Lifshitz transitions at the second Dirac magic angle, we show more density plots in the whole moir\'e Brillouin zone and zoomed in panels around the second magic angle. We also give the results of the AA/AB tetralayer and AA/ABC pentalayer graphene. All of them share the same Dirac magic angles.
\end{abstract}

\date{\today}

{
\let\clearpage\relax
\maketitle
}

\date{\today}

\onecolumngrid

\section{Band spectra at the origin of Dirac magic angles.}
\begin{figure}[ht]
   \centering
   \includegraphics[width=5in]{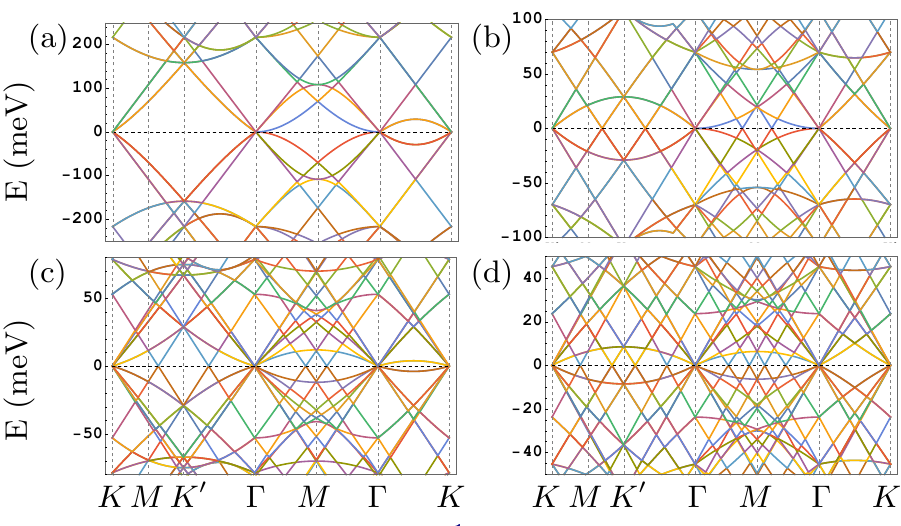} 
   \caption{Band spectra at $w=0$, $u=0.816$ and $
   \gamma_{tw}=0$. (a) $\theta=1.224^{\circ}$. (b) $\theta=(1.224/2)^{\circ}$. (c) $\theta=(1.224/\sqrt{7})^{\circ}$.(d) $\theta=(1.224/\sqrt{13})^{\circ}$.}
   \label{fig:Bands1}
\end{figure}

\noindent
In this section, we show the band spectra at the origin of Dirac magic angles, i.e., $w=0$. In these band spectra, multiple bands are degenerate at the $K$ and $\Gamma$ points at the same time, see Fig.~\ref{fig:Bands1}. Since $w=0$, the single layer graphene is totally decoupled from the AA-stacked bilayer graphene, the band spectra are the results of the geometric structure of the moir\'e reciprocal lattice. In Fig.~\ref{fig:Bands1}, we set $u=0.816$ and $\gamma_{tw}=0$, particle-hole symmetry is preserved. As we increase $w$, particle-hole symmetry will be broken as a non-zero $u$ breaks the chiral symmetry even when $\gamma_{tw}=0$.

\section{The first Dirac magic angle and Gamma magic angle.}
\noindent

\begin{figure}[H]
   \centering
   \includegraphics[width=6in]{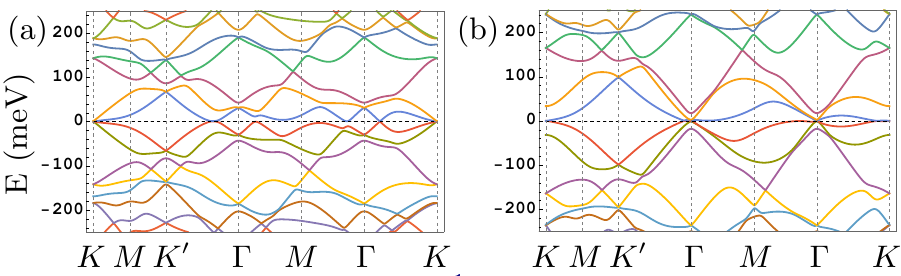} 
   \caption{Band spectra of (a) The first Dirac magic angle $\theta=1.31^{\circ}$. (b) The first Gamma magic angle $\theta=1.516^{\circ}$. The parameters $w=110$ meV, $u=0.816$ and $
   \gamma_{tw}=20$ meV.}
   \label{fig:Bands2}
\end{figure}

We plot band spectra of the first Dirac magic angle and the first Gamma magic angle, see Fig.~\ref{fig:Bands2}. The band spectrum of the first Dirac magic angle is similar to the second Dirac magic angle but without a clear gap between the positive bands and negative bands. The first Gamma magic angle is even larger than the first Dirac magic angle.

\section{The second Dirac magic angle}
\begin{figure}[H]
   \centering
   \includegraphics[width=4in]{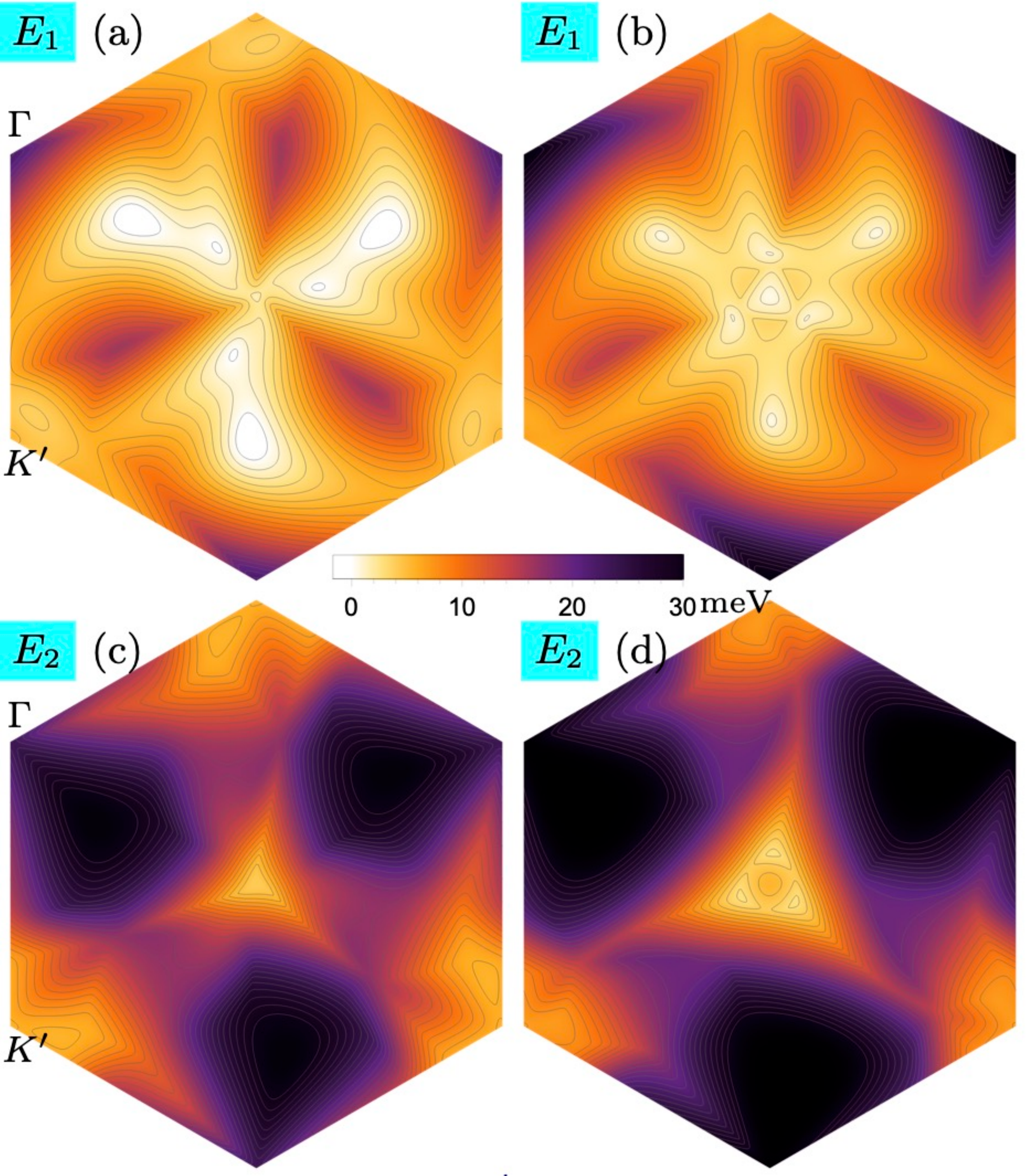} 
   \caption{Energy density plots of first positive bands (top row) and second positive bands (bottom row). (a) and (c) are at the Dirac magic angle $\theta=0.826^{\circ}$ with an electric field $\Delta V= 72.75$ meV. (b) and (d) are at the twisted angle $\theta=0.85^{\circ}$.}
   \label{fig:Bands3}
\end{figure}

\begin{figure}[H]
   \centering
   \includegraphics[width=4in]{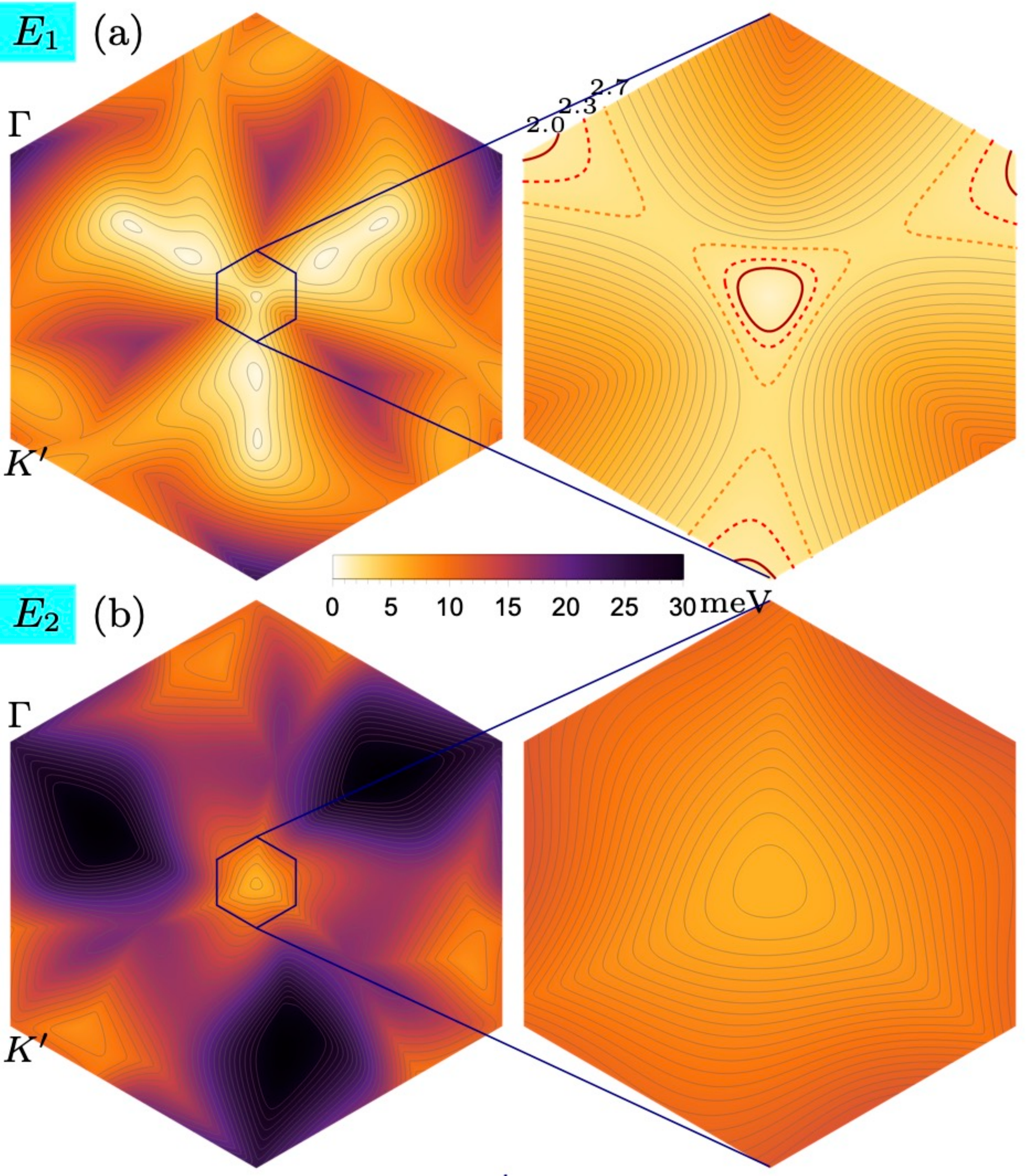} 
   \caption{Energy density plots at the angle $\theta=0.8^{\circ}$. (a) first positive bands (top row) with zoomed in panel and (b) second positive bands (bottom row) with zoomed in panel.}
   \label{fig:Bands4}
\end{figure}

\noindent
Since we focus on the second Dirac magic angle in the main text, we present here some more informations about it. In the main text, we show the zoomed in energy plots with the electric field $\Delta V =72.75$ and the angle $\theta=0.85^{\circ}$. Here we show the energy density plots in the whole moir\'e Brillouin zone, see Fig.~\ref{fig:Bands3}. We also show the density plots at the angle $\theta=0.8^{\circ}$ which is lower than the magic angle $\theta=0.826^{\circ}$, see Fig.~\ref{fig:Bands4}. We can see clearly that both electric field and twist angle changes can tune the Lifshitz transitions.

\section{The AA/AB tetralayer Dirac magic}
\noindent
The Hamiltonian of twisted AA/AB tetralayer graphene is
\begin{equation}
H_\text{AA/AB}=
\begin{bmatrix}
h_{\theta/2} & T_\text{AA} & 0 & 0\\
T_\text{AA}^\dagger & h_{\theta/2} & T & 0\\
0 & T^\dagger & h_{-\theta/2} & T_\text{AB}\\
0 & 0 & T_\text{AB}^\dagger & h_{-\theta/2}\\
\end{bmatrix},\label{eq:AAAB}
\end{equation}
where $T_\text{AA}=\gamma_\text{AA}$, $T_\text{AB}=\gamma_\text{AB}\cdot \sigma^{-}.$ We don't include the warping effect terms here and $\sigma^{-}$ is the lowering operator.

\begin{figure}[H]
   \centering
   \includegraphics[width=6in]{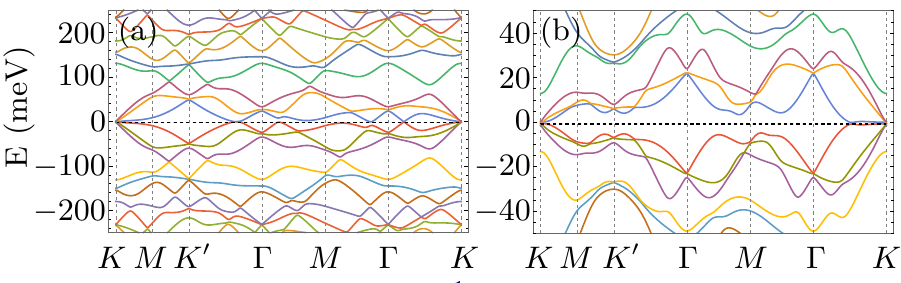} 
   \caption{Band spectra of AA/AB without warping (a) The first Dirac magic angle $\theta=1.31^{\circ}$. (b) The second Dirac magic angle $\theta=0.826^{\circ}$. The parameters $w=110$ meV, $u=0.816$ and $\gamma_{AB}=361$ meV~\cite{Jung_2014}.}
   \label{fig:Bands5}
\end{figure}

\begin{figure}[H]
   \centering
   \includegraphics[width=5in]{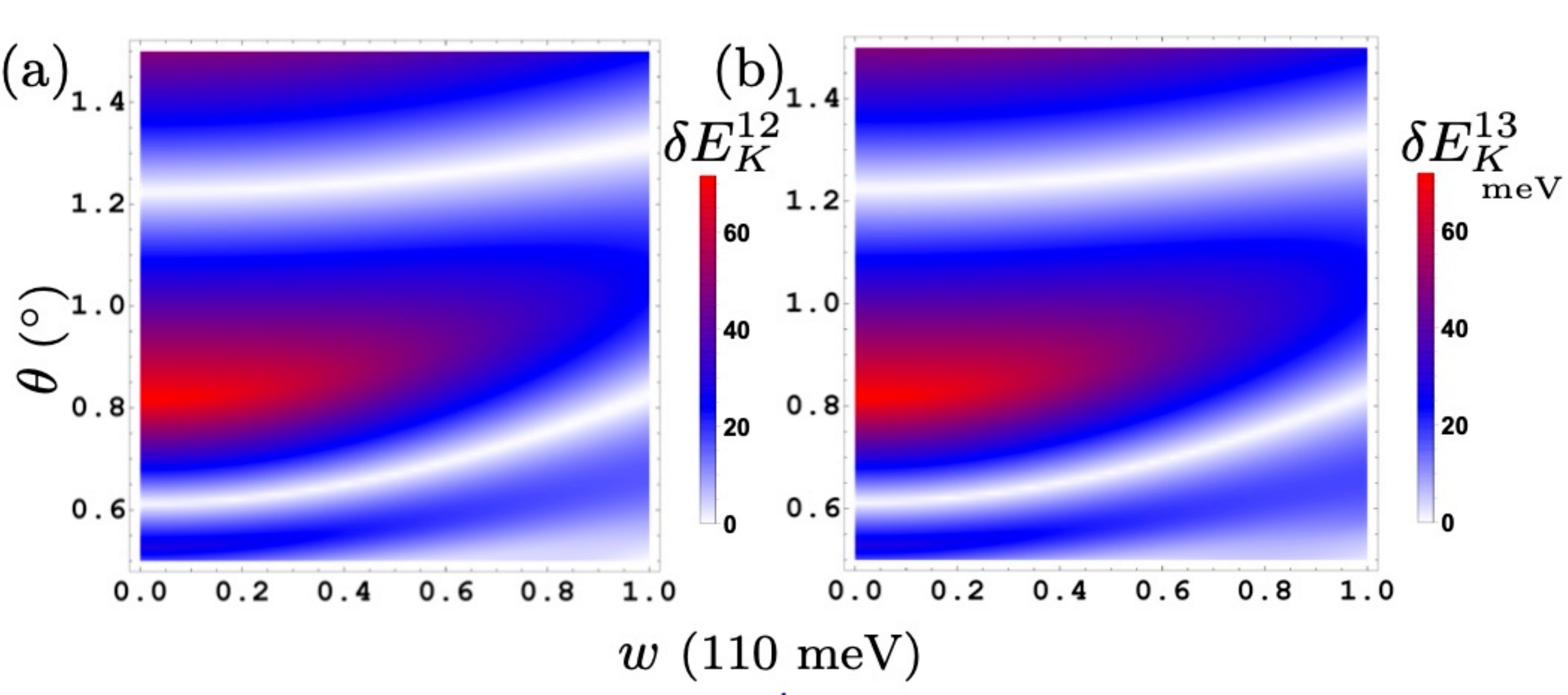} 
   \caption{(a) The energy splitting $\delta E^{12}_K = E_2-E_1$ at $K$ point for $u=0.816$. (b) The energy splitting $\delta E^{13}_K = E_3-E_1$ at $K$ point for $u=0.816$.}
   \label{fig:split6}
\end{figure}

We show the first and second Dirac magic angles in AA/AB tetralayer structure, see Fig.~\ref{fig:Bands5}. The magic angles are same as for the trilayer structure in the main text. The energy splittings at the $K$ point are shown in Fig.~\ref{fig:split6}. They are same for the second and third positive bands.

\section{The AA/ABC pentalayer Dirac magic}
\noindent
The Hamiltonian of twisted AA/ABC tetralayer graphene is
\begin{equation}
H_\text{AA/ABC}=
\begin{bmatrix}
h_{\theta/2} & T_\text{AA} & 0 & 0 & 0\\
T_\text{AA}^\dagger & h_{\theta/2} & T & 0 & 0\\
0 & T^\dagger & h_{-\theta/2} & T_\text{AB} & T_\text{AC}\\
0 & 0 & T_\text{AB}^\dagger & h_{-\theta/2} & T_\text{AB}\\
0 & 0 & T_\text{AC}^\dagger & T_\text{AB}^\dagger & h_{-\theta/2}\\
\end{bmatrix},\label{eq:AAABC}
\end{equation}
where $T_\text{AA}$ and $T_\text{AB}$ are same as we used in the AA/AB structure. $T_\text{AC}=\gamma_\text{AC}\cdot \sigma^{+}.$  Trigonal warping terms are not included here, and $\sigma^{+}$ is the raising operator.

\begin{figure}[H]
   \centering
   \includegraphics[width=6in]{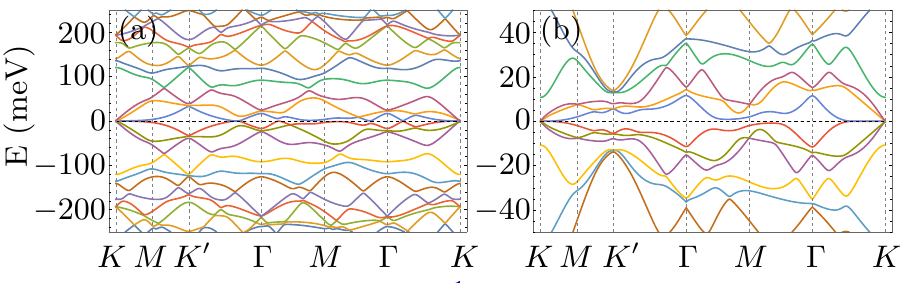} 
   \caption{Band spectra of AA/ABC without warping (a) The first Dirac magic angle $\theta=1.31^{\circ}$. (b) The second Dirac magic angle $\theta=0.826^{\circ}$. The parameters $w=110$ meV, $u=0.816$, $\gamma_{AB}=361$ meV and $\gamma_{AC}=0$.}
   \label{fig:Bands7}
\end{figure}

We show the first and second Dirac magic angles in AA/ABC pentalayer structure, see Fig.~\ref{fig:Bands7}. The magic angles are also same as the trilayer structure in the main text. Note that the two central bands become increasingly flat as we increase the number of layers, going from from trilayer to pentalayer.

\end{document}